\documentclass[structabstract]{aa}  
\usepackage{graphicx}
\usepackage{txfonts}
\usepackage{multirow}
\usepackage{longtable,lscape}
\usepackage[]{natbib}
\bibpunct{(}{)}{;}{a}{}{,}
\usepackage{stfloats}
\usepackage{fixltx2e}
\begin{document}
    \title{GRB 130606A within a sub-DLA at redshift 5.91}

%   \subtitle{I. Not in this case}

\author{A. J. Castro-Tirado \inst{1,2} 
 \and
  R. S\'anchez-Ram\'{\i}rez \inst{1} 
 \and 
  S. L. Ellison \inst{3} 
  \and 
  M. Jel\'inek \inst{1} 
   \and 
   A. Mart\'{\i}n-Carrillo \inst{4} 
 \and 
 V. Bromm \inst{5} 
 \and 
 J. Gorosabel \inst{1,6,7} 
 \and 
 M. Bremer \inst{8} 
 \and 
 J. M. Winters \inst{8} 
 \and 
 L. Hanlon \inst{4} 
 \and 
 S. Meegan \inst{4} 
 \and 
 M. Topinka \inst{4} 
 \and 
 S. B. Pandey \inst{9} 
 \and 
 S. Guziy \inst{10,1} 
 \and 
 S. Jeong \inst{1} 
 \and 
 E. Sonbas \inst{11} 
 \and 
 A. S. Pozanenko \inst{12} 
 \and 
 R. Cunniffe \inst{1} 
 \and 
 R. Fern\'andez-Mu\~noz \inst{13} 
 \and 
 P. Ferrero \inst{1} 
 \and 
 N. Gehrels \inst{14} 
 \and 
 R. Hudec \inst{15,16} 
 \and 
 P. Kub\'anek \inst{17} 
 \and 
 O. Lara-Gil \inst{1} 
 \and 
 V. F. Mu\~noz-Mart\'inez \inst{2} 
 \and 
 D. P\'erez-Ram\'irez \inst{18} 
 \and 
 J. \v{S}trobl \inst{15,16} 
 \and 
 C. \'Alvarez-Iglesias \inst{19,20} 
 \and 
 R. Inasaridze \inst{21} 
 \and 
 V. Rumyantsev \inst{22} 
 \and 
 A. Volnova \inst{12} 
 \and 
 S. Hellmich \inst{23} 
 \and 
 S. Mottola \inst{23} 
 \and 
 J. M. Castro Cer\'on \inst{24} 
 \and 
 J. Cepa \inst{19,20} 
 \and 
 E. G\"o\u{g}\"u\c{s} \inst{25} 
 \and 
 T. G\"uver \inst{26} 
 \and 
 \"O. \"Onal Ta\c{s} \inst{26} 
 \and 
 I. H. Park \inst{27} 
 \and 
 L. Sabau-Graziati \inst{28} 
 \and 
 A. Tejero  \inst{19,20}
}

\institute{
Instituto de Astrof\'{\i}sica de Andaluc\'{\i}a (IAA-CSIC), Glorieta de la Astronom\'{\i}a s/n, E-18008, Granada, Spain\\
\email{ajct@iaa.es}
\and
Unidad Asociada Departamento de Ingenier\'{\i}a de Sistemas y Autom\'atica, E.T.S. de Ingenieros Industriales, Universidad de M\'alaga, Spain.
\and 
University of Victoria, Department of Physics and Astronomy, P.O. Box 1700 STN CSC, Victoria, BC, V8W 2Y2, Canada.
\and
UCD School of Physics, University College Dublin, Belfield, Dublin 4, Ireland.
\and
Department of Astronomy, University of Texas, 2511 Speedway, Austin, TX 78712, USA.
\and
Unidad Asociada Grupo Ciencias Planetarias UPV/EHU-IAA/CSIC, Departamento de F\'{\i}sica Aplicada I, E.T.S., Ingenier\'{\i}a, Universidad del Pa\'{\i}s Vasco UPV/EHU, Bilbao, Spain.
\and
Ikerbasque, Basque Foundation for Science, Bilbao, Spain.
\and
Institute de Radioastronomie Millimétrique (IRAM), 300 rue de la Piscine, 38406 Saint Martin d' H\`eres, France.
\and
Aryabhatta Research Institute of Observational Sciences, Manora Peak, Nainital - 263 002, India.
\and
Nikolaev National University, Nikolska 24, Nikolaev, 54030, Ukraine.
\and
Department of Physics, University of Adiyaman, 02040 Adiyaman, Turkey.
\and 
Space Research Institute of RAS, Profsoyuznaya, 84/32, Moscow 117997, Russia.
\and
Instituto de Hortofruticultura Subtropical y Mediterr\'anea "La Mayora", Universidad de M\'alaga - Consejo Superior de Investigaciones Cient\'{\i}ficas (IHSM, UMA-CSIC), E-29750 Algarrobo-Costa (M\'alaga), Spain.
\and
NASA/GSFC, 8800 Greenbelt Road, Code 661, Bldg 34, Rm S254, Greenbelt, Maryland 20771, USA.
\and
Astronomical Institute, Academy of Sciences of the Czech Republic, Ond\u{r}ejov, Czech Republic.
\and
Czech Technical University in Prague, Faculty of Electrical Engineering, Prague, Czech Republic.
\and
Fyzik\'aln\'i \'ustav AV \u{C}R, v. v. i. Na Slovance 1999/2, 182 21 Praha 8, Czech Republic.
\and
Facultad de Ciencias Experimentales, Universidad de Ja\'en, Campus Las Lagunillas, E-23071  Ja\'en , Spain.
\and 
Instituto de Astrof\'{\i}sica de Canarias (IAC), C/. Via L\'actea s/n, E-38205 La Laguna (Tenerife), Spain.    
\and 
Departamento de Astrof\'{\i}sica, Universidad de La Laguna, C/. Molinos de Agua s/n, E-38200, Tenerife, Spain. 
\and 
Abastumani Astrophysical Observatory of Ilia State University G. Tsereteli Street 3, Tbilisi 0162, Georgia Republic.    
\and 
SRI “Crimean Astrophysical Observatory”, 98409,  Crimea, Nauchny, Ukraine.    
\and 
Institute of Planetary Research, DLR, Rutherfordstrasse 2, 12489 Berlin, Germany.    
\and 
European Space Astronomy Centre (ESA-ESAC), Camino bajo del Castillo, s/n, Villafranca del Castillo, E-28.692 Villanueva de la Ca\~nada (Madrid), Spain.
\and 
 Sabancı University, Orhanl{\i} - Tuzla, \.{I}stanbul 34956, Turkey.   
\and 
\.{I}stanbul University, Faculty of Science, Department of Astronomy and Space Sciences, 34119 University, \.{I}stanbul, Turkey.  
\and 
Department of Physics, Sungkyunkwan University, Suwon, Korea. 
\and 
Instituto Nacional de T\'ecnica Aerospacial (INTA), Ctra. de Ajalvir, km. 4, E-28850 Torrej\'on de Ardoz, Spain.    
}
\date{Received Dec XX, 2013; accepted Month Day, 2014}

% \abstract{}{}{}{}{} 
% 5 {} token are mandatory
 
\abstract
% context heading (optional)
% {} leave it empty if necessary  
{Events such as GRB\,130606A at z = 5.91, offer an exciting new window into pre-galactic metal enrichment in these very high redshift host galaxies.}
% aims heading (mandatory)
% {What do we want to do with this paper?}
{We study the environment and host galaxy of  GRB 130606A, a high-z event, in the context of a high redshift population of GRBs.}
% methods heading (mandatory)
%{How did we do it?}
{We have obtained multiwavelength observations from radio to gamma-ray, concentrating particularly on the X-ray evolution as well as the optical photometric and spectroscopic data analysis.}
% results heading (mandatory)
%{What came out of it?}
{With an initial Lorentz bulk factor in the range $\Gamma_0$ $\sim$ 65-220, the X-ray afterglow evolution can be explained by a time-dependent photoionization of the local circumburst medium, within a compact and dense environment. The host galaxy is a sub-DLA (log N (H I) = 19.85 $\pm$ 0.15), with a metallicity 
content  in the range from $\sim$1/7 to $\sim$1/60 of solar. Highly ionized species (N V and Si IV) are also detected.}
% conclusions heading (optional), leave it empty if necessary 
 {This is the second highest redshift burst with a measured GRB-DLA metallicity and only the third GRB absorber with sub-DLA HI column density. 
GRB ' lighthouses' therefore offer enormous potential as backlighting sources to probe the ionization and metal enrichment state of the IGM at very high redshifts for the chemical signature of the first generation of stars.}
\keywords{gamma-ray burst:general
}

\authorrunning{Castro-Tirado et al.}
\titlerunning{GRB 130606A within a sub-DLA at redshift 5.91}
\maketitle
%
%________________________________________________________________

\section{Introduction}
It has been recently suggested \citep{Cooke:2011aa} that very metal-poor damped Lyman-alpha (DLA) systems (regions of high column density of neutral gas at high redshifts \citep{Wolfe:aa}) could bear the chemical signature of the first generation of stars (Population III stars, \cite{Bromm:aa}) born a few hundred million years after the Big Bang. 
Indeed, it has been suggested that metal-free regions persist to values of z $\leq$ 6, allowing Pop III stars with masses in the range 140-260 solar masses to be observed as pair-production instability supernovae \citep{Scannapieco:2005aa}, although a core-collapse (Type II) supernova instead is also plausible \citep{Wang:2012aa}.
In spite of the fact that no GRB has so far been firmly associated with a Pop III collapse yet, the high z values found for several GRBs 
%Our observations can be interpreted as the result of the GRB radiation penetrating material in the host galaxy which was pre-enriched by these Population III stars with the first heavy chemical elements. This 
reinforce the potential of GRBs to provide bright background sources to illuminate the early intergalactic medium at a time when quasars were too rare and dim to serve this purpose.\

A $\sim$275 s cosmic gamma-ray burst (GRB\,130606A) was recorded by {\it Swift} and KONUS-{\it Wind} on 6 June 2013, 21:04:34 U.T. ($T_0$) \citep{Barthelmy_gcn2013, Golenetskii:2013aa}, displaying a bright afterglow (the emission at other wavelengths following the gamma-rays) in X-rays, but no apparent optical transient emission \citep{Ukwatta:2013aa} in the range of the UVOT telescope aboard {\it Swift}. The TELMA 0.6m diameter telescope at the BOOTES-2 station automatically responded to the alert and an optical counterpart was identified \citep{Jelinek:2013aa}, thanks to the spectral response of the detector up to 1 $\mu$m, longer than that of {\it Swift}/UVOT (0.17-0.65 $\mu$m). 

%Prompt optical observations of GRB\,130606A previous to the BOOTES-2/TELMA discovery were carried out by the 0.4m WATCHER telescope in South Africa starting on June 6, 21:06:49 U.T., i.e. $\sim$2.25 minutes after the initial {\it Swift}/BAT trigger, which we also report in this work, and complemented with the 0.25m BART and 1.23m CAHA observations.  

The detection of the afterglow at BOOTES-2/TELMA prompted spectroscopic observations with the 10.4\,m Gran Telescopio Canarias (GTC) starting 1.4 hr after the event, which revealed a very distant explosion at a very high redshift (z $\sim$ 6) \citep{Castro-Tirado:2013aa}, a value later refined to z = 5.9135 $\pm$ 0.0005 \citep{Castro-Tirado:2013ab}, when the Universe was only $\sim$950 million years old. 

In this work  we provide details (Section 2) of the optical imaging and spectroscopic observations beginning a few minutes after the onset of this distant explosion, as well as publicly available $\gamma$/X-ray data from the {\it Swift} satellite, complemented with additional millimetre observations. Results are presented in Section 3 and conclusions in Section 4. 

%with at least three other intervening systems (also partly reported in \cite{Chornock:2013aa}), in the line of sight, at redshifts z = 2.310, 2.521 and 3.451,  see Fig~\ref{fig:spec}. 

\section{Observations}
\label{Observations}

\subsection{Optical Observations}
%\indent
\label{OptObs}

\begin{table*}[H]
\caption{Reference stars in the field of GRB\,130606A.} 
\label{table:stdlog}
\begin{tabular}{cccccc}
\hline
\noalign{\smallskip}
N & R.A.(J2000) & Dec(J2000)  &  R-band mag  &  I-band mag & H-band mag  \\
\noalign{\smallskip}
\hline
\noalign{\smallskip}
1  & 16 37 33.7 & +29 48 19.0 &  18.04 $\pm$ 0.11  &  16.36 $\pm$ 0.02 & 14.54 $\pm$ 0.05 \\
2  & 16 37 28.4 & +29 47 05.6 &  16.97 $\pm$ 0.07  &  15.71 $\pm$ 0.02 & 13.95 $\pm$ 0.04 \\
3  & 16 37 31.9 & +29 46 53.6 &  16.41 $\pm$ 0.05  &  16.12 $\pm$ 0.02 & 15.31 $\pm$ 0.05 \\
4  & 16 37 39.4 & +29 49 05.6 &  16.22 $\pm$ 0.05  &  15.88 $\pm$ 0.02 & 15.07 $\pm$ 0.05 \\
5  & 16 37 31.0 & +29 49 36.4 &  17.65 $\pm$ 0.09  &  17.03 $\pm$ 0.03 & $--$ \\
6  & 16 37 40.4 & +29 48 03.4 &  18.62 $\pm$ 0.14  &  18.21 $\pm$ 0.04 & 17.38 $\pm$ 0.06\\
7  & 16 37 39.5 & +29 46 07.7 &  15.99 $\pm$ 0.04  &  15.66 $\pm$ 0.02 & $--$ \\
8  & 16 37 26.6 & +29 47 34.5 &  14.57 $\pm$ 0.02  &  14.24 $\pm$ 0.02 & 13.54 $\pm$ 0.04\\
9  & 16 37 44.8 & +29 48 25.3 &  13.11 $\pm$ 0.02  &  12.82 $\pm$ 0.02 & $--$ \\
\noalign{\smallskip}
\hline
\end{tabular}
\end{table*}

\subsubsection{Photometry}
\indent
\label{OptPhot}
\indent   Early time prompt optical observations were carried out by the Watcher telescope starting on June 6, 21:06:49 U.T., i.e. $\sim$\,135 s after the first {\it Swift}/BAT trigger (T0 = 21:04:34 UT). The BOOTES-2/TELMA observations, which resulted in the optical afterglow discovery, started on June 6,  21:17:33 U. T., i.e. $\sim$660 s after the first {\it Swift}/BAT trigger. The Watcher observations partially cover the second {\it Swift}/BAT peak. Additional Johnson R and V--band images were acquired with the 1.23m telescope of Calar Alto (CAHA) observatory, Spain. Late epoch optical observations were obtained with the 0.7m Abastumani Observatory, the AZT-11 (1.25m) telescope at SRI “Crimean Astrophysical Observatory”, the T100 (1m) telescope at T\"UBITAK National Observatory, the 1.5m OSN telescope at Observatorio de Sierra Nevada and with the 10.4m Gran Telescopio Canarias (GTC) equipped with the OSIRIS imaging spectrograph (Fig. 1). Optical photometry is based on isophotal corrected photometry by IRAF/PHOT31 against standard reference Landolt fields imaged at the 1.5m OSN telescope in order to provide reference stars in the field (Table 1). The photometric results of the afterglow are tabulated in Table~\ref{table:Optlog}.  

\subsubsection{Spectroscopy}
\indent
\label{OptSpe}
\indent Starting 1.3 hr post-burst, optical spectra were obtained on 6 June 2013 with the 10.4m GTC using the R1000B and R500R grisms (1\,$\times$\,450\,s exposures) and R2500I (2\,$\times\,$1,200\,s exposures) of the OSIRIS imaging spectrograph. The later one provides a nominal resolution of $\sim$ 120 km s$^{-1}$. The log is given in Table~\ref{table:speclog}.  The 1" wide slit was positioned on the location of the transient source and a 2 $\times$ 2 binning mode was used. The GTC spectra were reduced and calibrated following standard procedures using custom tools based on IRAF and Python. Standard spectrophotometric stars used for flux calibration were Feige 92 for observations for the R1000B grism and Ross 640 for the prisms R500R and R2500I, taken the same night.  All spectra were scaled in flux to correct for slit losses using the photometry of the corresponding acquisition images.  The final wavelength calibration is on a vacuum scale, appropriate for the application of rest-frame UV atomic data.

\subsection{Near-IR observations}
\indent
\label{NirObs}
\indent Near-IR bservations were conducted on July 22 at the 3.5m telescope (+\,OMEGA 2000) at the German-Spanish Calar Alto (CAHA) Observatory, with a 5,400s overall exposure time in the H-band. 
The photometric calibration is based on the observation of the standard S889-E 
 \citep{Person:1998aa} and the result is given in Table~\ref{table:Optlog} 
%(Persson et al. 1998, AJ 116, 2475) 
at an airmass similar to the GRB field.
%The photometric calibration against the reference stars is done accordingly. The upper limit at the position of the GRB host galaxy was determined from the 2MASS point source catalogue.

\begin{figure}[h]
\centering
   \includegraphics[width=9cm]{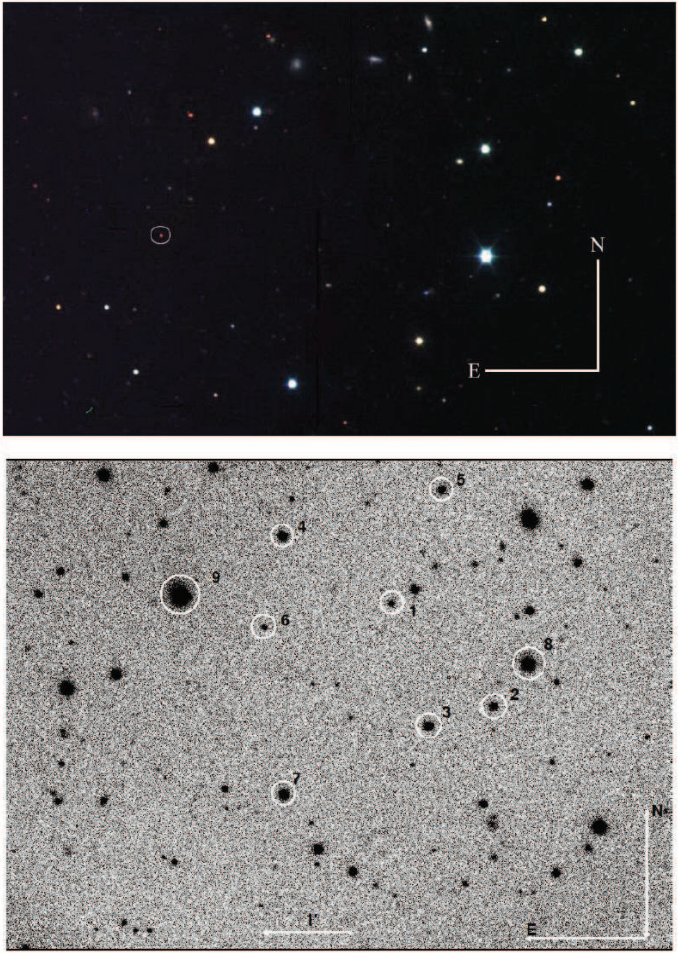}
\caption{The colour composite image of the field around GRB\,130606A and reference stars in the line of sight. Upper panel: The highly reddened GRB afterglow (circled) and the surrounding field, based on g’r’i’ images obtained at the 10.4m GTC on June 7, 2013. The field of view is 3.8 $\times$ 2.8 $\mathrm{{arc min}^{2}}$. Lower panel: Reference stars for photometric calibration in the field around GRB\,130606A (Table~1). The field of view (r’-band image) is 6.8 $\times$ 5.0 $\mathrm{{arc min}^{2}}$.}
\label{fig:pho}
\end{figure}

%----------------------------------------------------------- 

%----------------------------------------------------------- 

\begin{table*}[h]
\caption{Optical and near-IR observations gathered at several astronomical observatories worldwide. RVIH-band magnitudes are given in the Vega system whereas clear and  Sloan-filter magnitudes are given in the AB system. Not corrected for Galactic extinction.} 
\label{table:Optlog}
\centering
\begin{tabular}{cccc}
\hline
Start Time (JD) \tablefootmark{a}    &  Magnitude   &   Filter    &   Telescope\tablefootmark{b}\\
\hline\hline
2456450.379734  & 17.99 $\pm$ 0.11 &   clear  &     0.4m Watcher \\    
2456450.381019 &  17.55 $\pm$ 0.08 &   clear  &  \\
2456450.381273 &  17.76 $\pm$ 0.10 &   clear  &  \\
2456450.381528 &17.04 $\pm$ 0.06  & clear    &   \\
2456450.386308  &16.99 $\pm$ 0.10  &  clear    &  \\
2456450.391528 &17.38 $\pm$ 0.14  &    clear    & \\
\hline
2456450.382210  &  $>$ 16.5    &   R     &     0.25m BART \\
2456450.38575  &   18.30 $\pm$ 0.40   & R   &   \\
\hline
2456450.385800 & 16.73 $\pm$ 0.34  &  i’     &    0.6m TELMA \\
2456450.387112 & 17.16 $\pm$ 0.16  &  i’      & \\
2456450.388424 & 17.10 $\pm$ 0.13  &  i’     &  \\
2456450.389742 & 17.18 $\pm$ 0.12 &   i’    &  \\
2456450.391053 & 17.59 $\pm$ 0.20  &   i’    &  \\
2456450.392371 & 17.70 $\pm$ 0.20 &   i’    &\\
2456450.393677 & 17.32 $\pm$ 0.16 &    i’   & \\
2456450.395124 & 17.42 $\pm$ 0.16 &   i’    & \\
2456450.396963 &  17.56 $\pm$ 0.15 &   i’   &  \\
2456450.399062 &  17.38 $\pm$ 0.15  &   i’  & \\
2456450.401339 & 17.99 $\pm$ 0.21 &     i’  & \\
2456450.403521 & 18.04 $\pm$ 0.24 &    i’   &  \\
2456450.406057 & 18.29 $\pm$ 0.21 &    i’   &   \\
2456450.408959 & 18.61 $\pm$ 0.30  &    i’   &  \\
\hline
2456450.514739 & 17.49 $\pm$ 0.15   & Z   &  0.6m TELMA\\
2456450.537466 & 17.56 $\pm$ 0.16  &   Z   & \\
2456450.560270 &  17.93 $\pm$ 0.15  &  Z   &  \\
2456450.583792 & 17.84 $\pm$ 0.16  &  Z    & \\
2456450.606892 & 17.89 $\pm$ 0.17   &  Z     & \\
2456450.629280 & 18.32 $\pm$ 0.26   &  Z     &   \\
2456450.651610 & 18.37 $\pm$ 0.31   &   Z      & \\
\hline
2456450.392159 & 18.75$ \pm$ 0.03  &   R      &  1.23m CAHA   \\
2456450.395772 & 18.99 $\pm$ 0.04 &    R  & \\
2456450.399383 &  19.26 $\pm$ 0.06  &    R  & \\
\hline
2456450.403071  &   $>$21.5    &      V  &   1.23m CAHA    \\
\hline
2456450.408137 &   18.84 $\pm$ 0.05  & clear  & 0.7m AO  \\ 
2456450.410694  & 18.86 $\pm$ 0.04 & clear &  \\
2456450.413241 & 18.95 $\pm$ 0.05 & clear &   \\
2456450.415799 & 19.16 $\pm$ 0.06 &  clear  &  \\
\hline
2456450.416655 &  21.08 $\pm$ 0.38 &   R   &   1.25m AZT-11   \\       
2456450.447046 &  20.58 $\pm$ 0.07   & R    &  1.0m T100    \\
2456450.451171 & 20.71 $\pm$ 0.07 &    R  &  \\
2456450.455296 &  20.51 $\pm$ 0.07 &   R  &   \\
2456450.459379 &  20.60 $\pm$ 0.08  &   R  &   \\
2456450.463463 &  20.86 $\pm$ 0.09  &    R  &    \\
\hline
2456450.582700  &   $>$25    &      g’    &       10.4m GTC  \\
\hline
2456450.583935   &   21.73 $\pm$ 0.07  &    r’   & 10.4m GTC\\
2456450.120092  &   22.00 $\pm$ 0.07  &    r’   &    \\
2456450.429560  &  19.27 $\pm$ 0.05  &    i’     &    10.4m GTC\\
2456450.584792  &  20.53 $\pm$ 0.05  &    i’       &   \\ 
2456450.621551  &  20.83 $\pm$ 0.05   &     i’     &  \\      
\hline     
2456450.585938  &  18.14 $\pm$ 0.08  &   z’      &  10.4m GTC\\
2456450.622558  &   18.42 $\pm$ 0.08   &   z’     &  \\
\hline
2456451.480060  &   $>$22.1  &      I     &      1.5m OSN   \\              
2456496.462500  &  $>$21.5    &   H       &    3.5m CAHA  \\ 
\hline
\end{tabular}
\tablefoot{
\tablefoottext{a}{Values measured since 21:04:34 UT June 6, epoch of the first {\it Swift}/BAT trigger time (Julian Date (JD) is 2556450.378171)}\\
\tablefoottext{b}{0.4m Watcher is the 0.4m telescope in Boyden Observatory (South Africa). 0.25m BART is the 0.25m telescope in Astronomical Institute at Ond\u{r}ejov (Czech Republic). 0.6m TELMA is the 0.6m TELescopio Malaga in Algarrobo Costa (M\'alaga, Spain). 0.7m AO is the 0.7m AS-32 telescope in Abastumani Observatory (Georgia).  T100 is the 1.0m telescope in T\"UBITAK National Observatory (Turkey). AZT-11 is the 1.25m telecope at the SRI “Crimean Astrophysical Observatory” (Ukraine). 1.5m OSN is the 1.5m telescope at Observatorio de Sierra Nevada in Granada (Spain); 3.5m CAHA  is the 3.5m telescope at the German-Spanish Calar Alto Observatory in Almer\'ia (Spain); GTC is the 10.4-m Gran Telescopio Canarias in Canary Islands (Spain).}
}
\end{table*}

%__________________________________________________ 

\begin{table*}[H]
\caption{Log of Spectroscopic data obtained at the 10.4m GTC.} 
\label{table:speclog}
\begin{tabular}{cccccc}
\hline
\noalign{\smallskip}
Start Time (UT)  &  Exp Time (s)  &   Grism    &   Wavelength Range (\AA) & Slit width (") & Airmass \\
\noalign{\smallskip}
\hline\hline
\noalign{\smallskip}
06-Jun 2013-22:23:50.5 &   1  x  450 &     R1000B  &         3,650 -  7,750 &  1.2 &        1.17 \\
06-Jun 2013-22:32:18.1  &  1  x  450     &   R500R      &    4,750 - 10,300     &   1.2  &  1.15 \\ 
07-Jun 2013-02:10:09.7  &  2  x 1,200 &    R2500I   &       7,320 - 10,100    &      1.0     &   1.05 \\
\noalign{\smallskip}
\hline
\end{tabular}
\end{table*}

       %__________________________________________________

\begin{table*}[H]
\caption{Flux densities measured at Plateau de Bure Interferometer.} 
\label{table:radlog}
\begin{tabular}{ccc}
\hline
\noalign{\smallskip}
Time (days post-burst)      &  Flux density at 3 mm [mJy] &  Frequency [GHz] \\
\noalign{\smallskip}
\hline
\noalign{\smallskip}
 3.30       &    1.45 $\pm$ 0.15   &   86.7 \\
 7.50       &    0.03 $\pm$ 0.13    &  86.7 \\
\noalign{\smallskip}
\hline
\end{tabular}
\end{table*}

 %__________________________________________________ 

\begin{figure}
\centering
   \includegraphics[width=9cm]{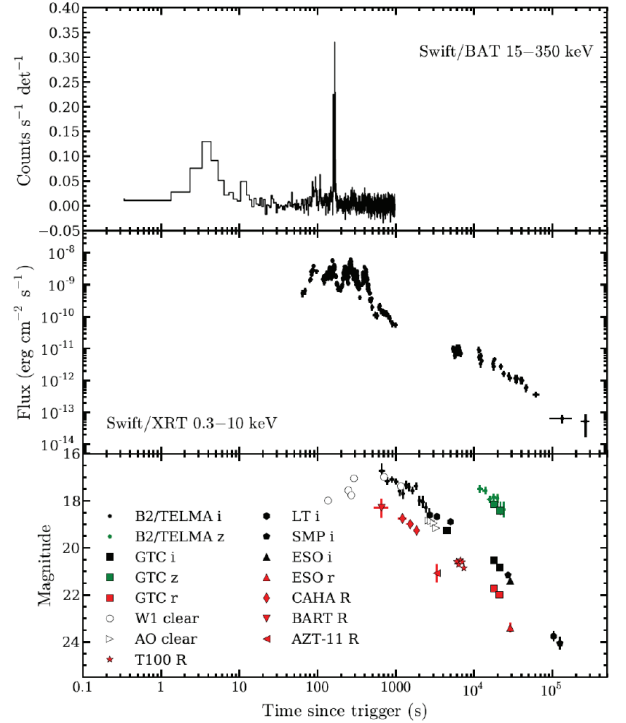}
\caption{The GRB 130606A  prompt gamma-ray emission and the multiwavelength afterglow evolution.  The {\it Swift}/BAT light curve shows a double-peaked
structure with the initial peak lasting $\sim$10 s and a brighter second 
peak at T$_{0}$+150 s of $\sim$ 20 s duration. The gamma-ray lightcurve is compared with the multiwavelength (X-ray, optical) GRB 130606A afterglow lightcurves. Significant temporal (and spectral) evolution is noticeable in the XRT data. The lower panel shows the rising optical afterglow lightcurve based on Watcher data, prior to the well sampled decay, based on the data gathered by BART, BOOTES-2/TELMA, 0.7m AO, T100, 1.23m CAHA, AZT-11, 1.5m OSN and 10.4m GTC (Table 1), complemented with other data published elsewhere  \citep{Afonso:2013aa,Butler:2013aa,Virgili:2013aa}. The observations were carried out in different optical wavebands (R, r , i, z) or clear (meaning unfiltered) and the initial bulk Lorentz factor to be determined. 1$\sigma$ error bars are plotted.} 
\label{fig:lc}
\end{figure}

\begin{figure}[t]
\centering
   \includegraphics[width=8cm]{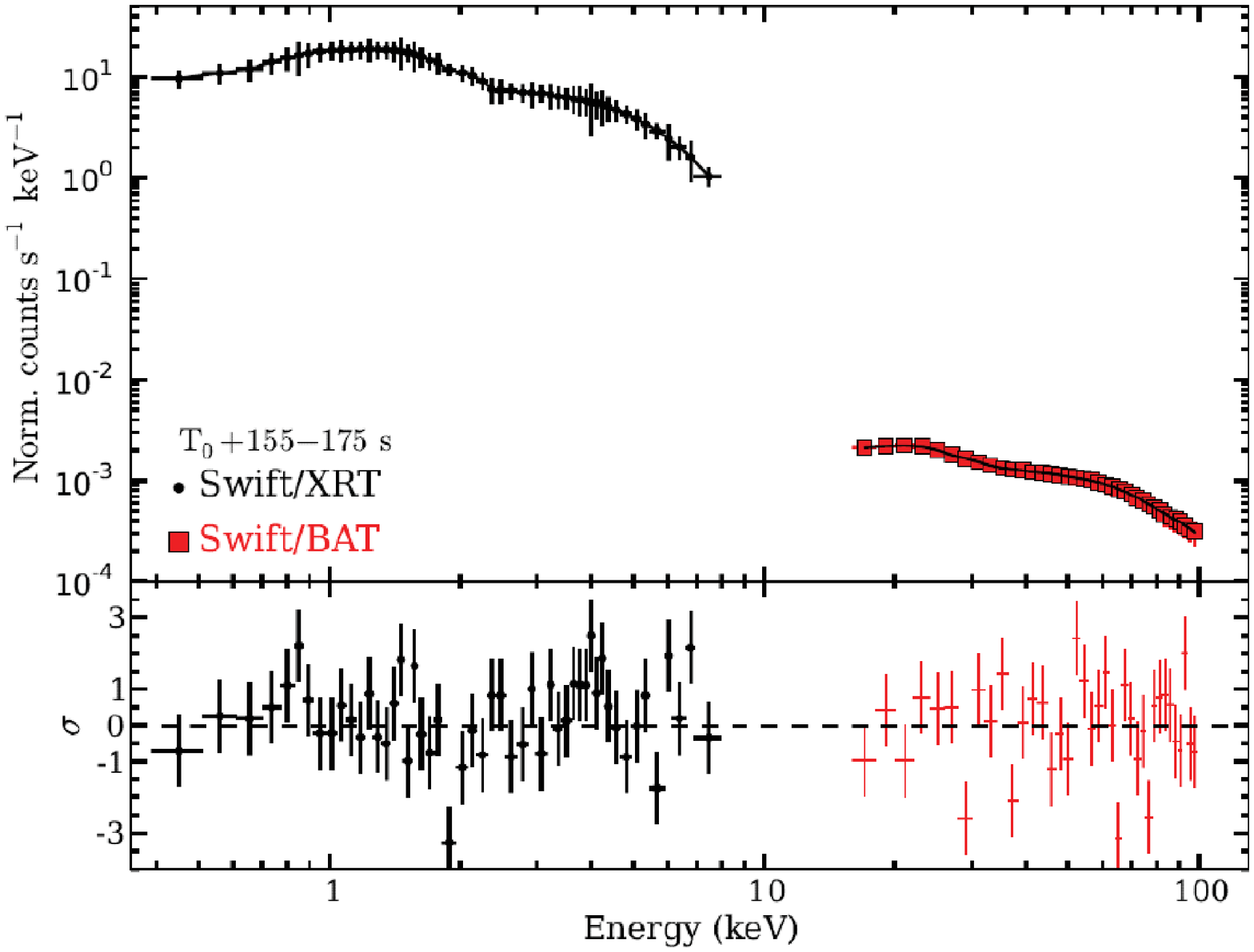}
\caption{The GRB\,130606A {\it Swift}/BAT and {\it Swift}/XRT spectrum. A simple power-law function (black line) yields a formally acceptable fit ($\mathrm{\chi^{2}}$ /d.o.f. = 1.29). The inclusion of a thermal component provides negligible improvement ($\mathrm{\chi^{2}}$ /d.o.f. = 1.26). 1$\sigma$ error bars are shown.} 
\label{fig:bat+xrtspec}
\end{figure}

\subsection{Millimetre observations}
\indent
\label{MilObs}
\indent 
Target-of-Opportunity millimetre observations were carried out at the Plateau de Bure Interferometer (PdBI) \citep{Guilloteau:1992aa}. It was pointed to the GRB 130606A location on two occasions in its compact 6 antenna configuration. The millimetre counterpart was detected 3.30 days after the GRB onset with a high ($\sim$10) S/N ratio, on the phase center coordinates (J2000, R.A. = RA: 16:37:35.13; Dec: +29:47:46.5). The results of UV-plane point source fits to the phase center are given in Table~\ref{table:radlog}.

The millimetre afterglow was detected with a flux density  of $\sim$1.5 mJy at 3 mm, confirming the detection earlier reported at centimetre wavelengths \citep{Laskar:2013aa}. The source became undetectable by June 14.

\begin{figure}
\centering
   \includegraphics[width=8cm]{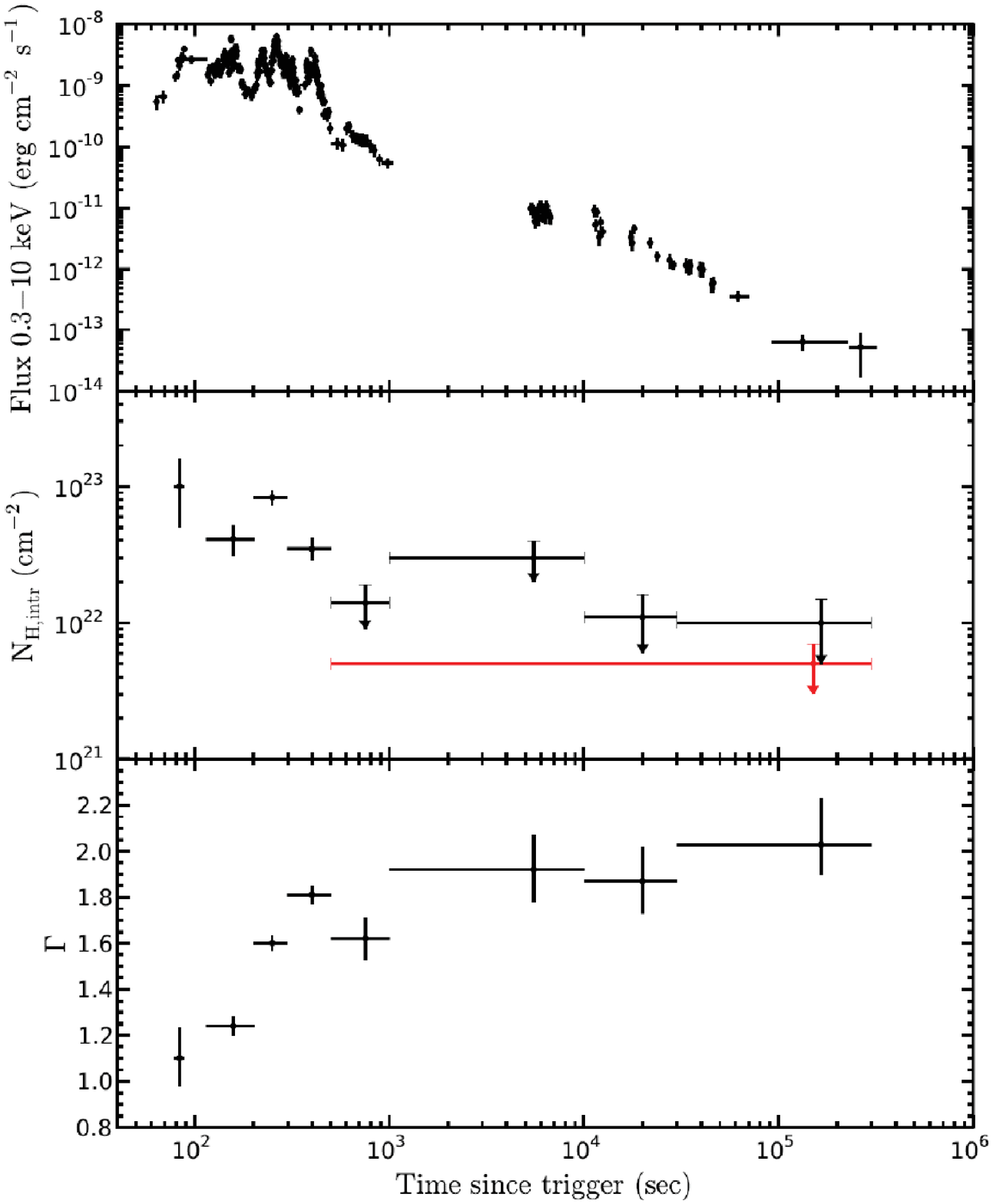}
\caption{X-ray light curve for the GRB\,130606A afterglow and variations in the power-law index and column density as a function of time. Upper panel: The X-ray afterglow light curve. Middle panel: The decrease of the column density as the gamma/X-ray emission decreases. Lower panel: The variation of the power-law index $\Gamma$ with respect to the X-ray luminosity showing that the spectrum gets progressively harder as the flux increases. 1$\sigma$ error bars are shown. } 
\label{fig:xrtlc}
\end{figure}

\begin{figure}
\centering
   \includegraphics[width=8cm]{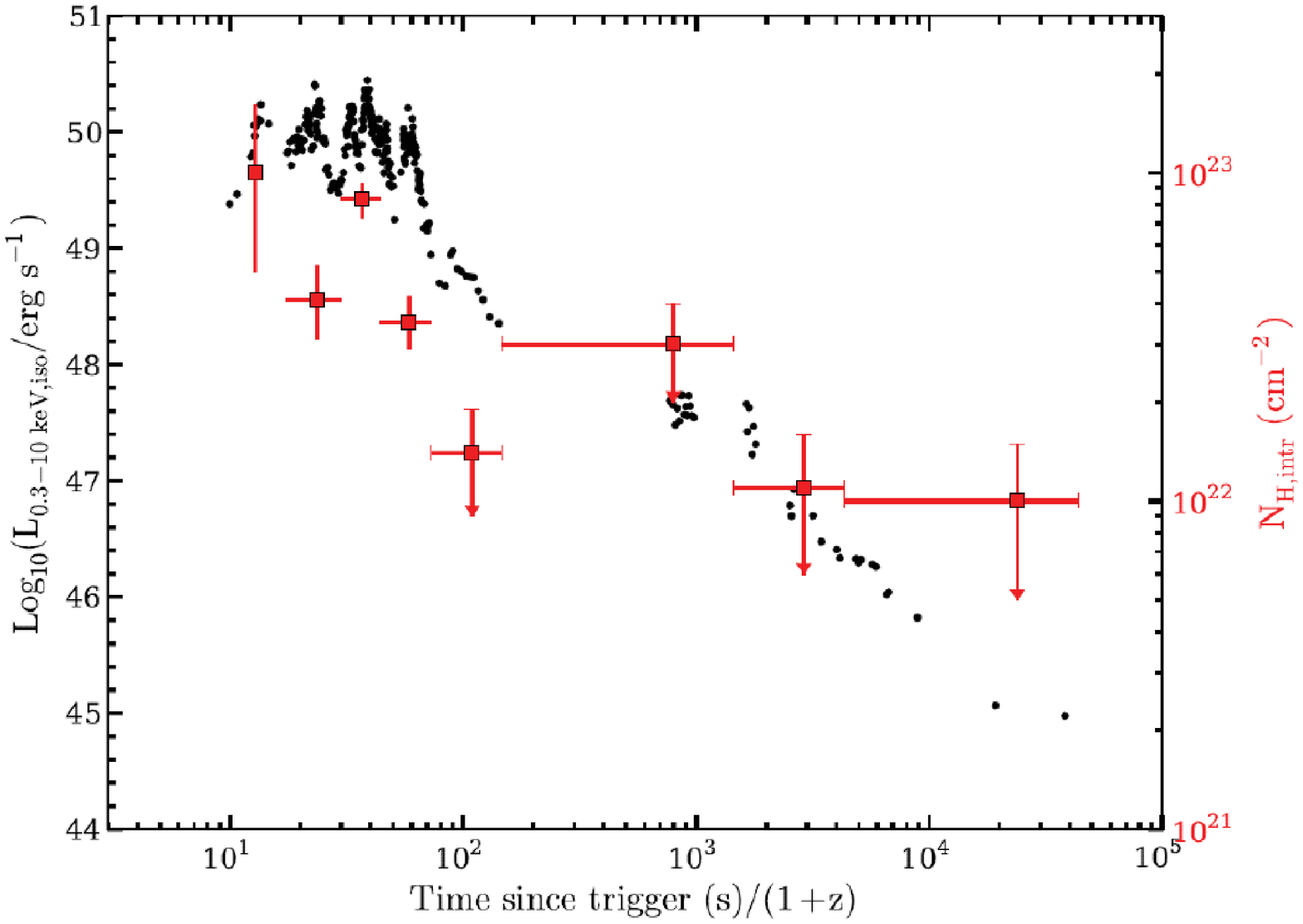}
\caption{The GRB\,130606A X-ray afterglow and variations in the power-law index and column density. The variation of the $N_H$ column density (squares with red bars) with respect to the X-ray luminosity (black dots). 1$\sigma$ error bars are shown.} 
\label{fig:nh}
\end{figure}

\subsection{X-ray observations}
\indent
\label{XObs}
\indent The {\it Swift} X-ray Telescope (XRT) started collecting data in window timing mode (WT) $\sim$ 60s after the initial BAT trigger on June 6 at 21:05:35 UT, switching to photon counting mode (PC) after $\sim$ 500s. The position of the source was monitored up to $\sim$ 3 $\times$ $\mathrm{10^{5}}$ s post-trigger. The data were processed using the {\it Swift} software v.2.6.  A cleaned event file was generated using the default pipeline, which removes the effects of hot pixels and Earth brightness. From this cleaned event list, the source and background light curves and spectra were extracted from a region of 20” using xselect. In general, no pile-up was found in either the WT or the PC data. The data have been fit using a fixed Galactic column density at 2 $\times$ $\mathrm{10^{20} \,cm^{-2}}$  and a varying column density in the rest frame in the range 2-6 $\times$ $\mathrm{10^{22} \,cm^{-2}}$.

\section{Results}
\label{Results}

Hereafter we consider $H_0$ = 73 km $s^{-1}$ $Mpc^{-1}$, $\Omega_\Lambda$= 0.73   $\Omega_m$  = 0.27.  At the redshift of z = 5.9135 (see below), the light travel time was 12,350 Gyr, the age of the Universe at this redshift was 0.95 Gyr and the luminosity distance is 56,365 Mpc.
 \subsection{The initial bulk Lorentz factor}
 \indent  The 0.6m BOOTES-2/TELMA and 1.23m CAHA colours show clear signs of a high-redshift dropout (V-R$>$2.2 in the CAHA case).
From the optical light curve depicted in Fig.~\ref{fig:lc}, the optical flux ($F_{opt}$ $\propto$ $\mathrm{t^{\alpha}}$) exhibits a  rising phase toward an optical maximum at $\sim$ 7.5 minutes after the BAT trigger. A  rising temporal index $\alpha_1$ $\sim$ 1.2 and decaying temporal index of $\alpha_2$ $\sim$ -1.25 are derived, with a break around t $\sim$ 450 s. The interstellar medium (ISM) case predicts the rising index to be $\sim$ 2 whereas a wind profile (WIND) case predicts it to be $\sim$ 0.5 \citep{Panaitescu:2011aa}. These values have not been seen in many of the observed cases at early times, possibly due to the early emission being a combination of multiple components such as early time energy injection. Using Eq. 4 of \cite{Rykoff:2009aa} and for s $\sim$ 4, the peak time $t_{peak}$ is $\sim$ 350 s. Assuming that the early optical emission is the onset of the forward shock emission, the value of $t_{peak}$ in the rest frame can be used (i.e. $t_{peak}$/(1+z)) to calculate the initial bulk Lorentz factors for the ISM and WIND cases for the GRB environment, following \cite{Melandri:2010aa}. We also consider the isotropic energy released in GRB\,130606A (at z $\sim$ 5.91) to be 28.3 $\pm$ 0.5 in units of $\mathrm{10^{52}}$ ergs \citep{Golenetskii:2013aa}. Thus, for $E_{52}$ $\sim$ 28.3 and $t_{peak}$ $\sim$ 350 s, the bulk Lorentz factor in the ISM case is $\Gamma_0$ $\sim$ 185, whereas in the case of WIND, $\Gamma_0$ $\sim$ 65. $\Gamma_0$ can be also estimated from \cite{Ghirlanda:2012aa}, giving  $\Gamma_0$  $\sim$ 160 for the ISM case and $\Gamma_0$ $\sim$ 70 for the WIND case. Using  the statistical relations from \cite{Liang:2010aa} and  \cite{Lu:2012aa}, we find $\Gamma_0$ $\sim$ 185 and $\sim$ 220 respectively for the ISM and WIND cases. The post-peak data of the optical light curve exhibits a temporal decay index
$\alpha_{opt}$ $\sim$ 1.3 along with a superimposed variability.

\begin{figure*}[t]
\centering
   \includegraphics[width=19cm]{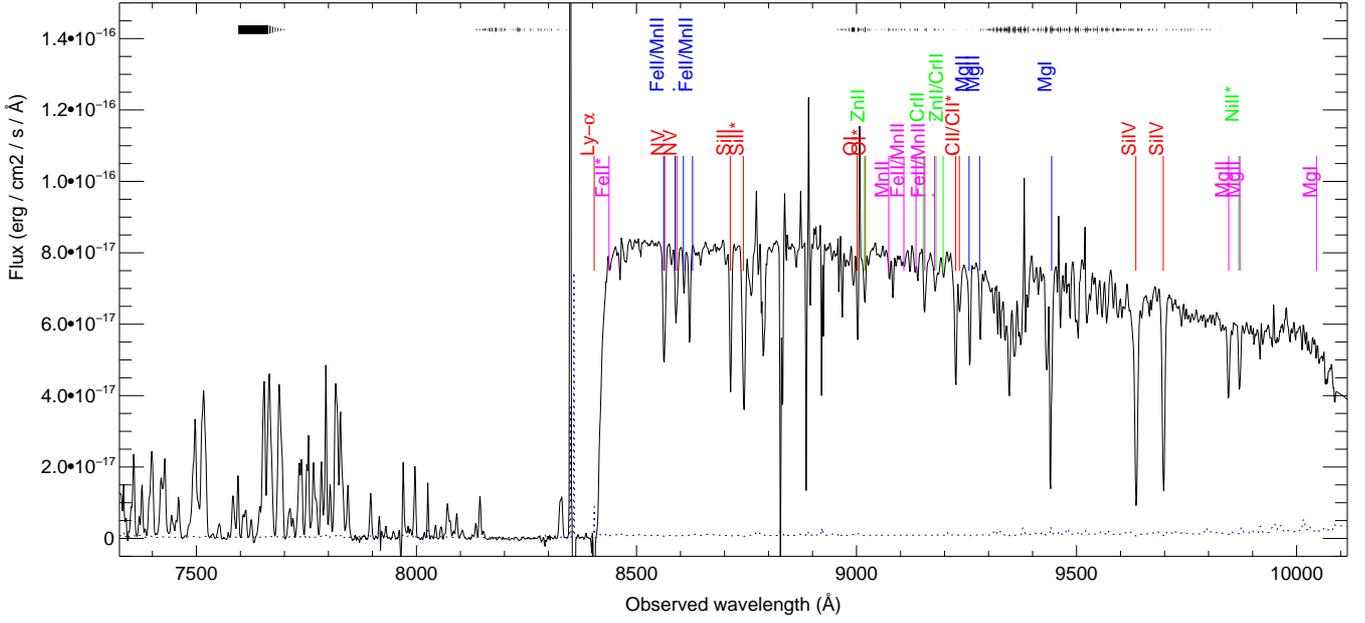}
\caption{The 10.4m GTC spectrum $\sim$ 6 hr post-burst. This is the highest resolution spectrum taken for GRB 130606A by GTC, with the R2500I grism (see Table 3 for details). The noise spectrum (dotted line) is also plotted. Intervening systems at redshifts z = 2.310 (blue), 2.521 (magenta), 3.451 (green) besides that (a sub-DLA) of the GRB 130606A host galaxy at z = 5.913 (red) are plotted. Strong absorption by intergalactic hydrogen in the line of sight is causing the apparent low optical flux observed in the Lyman-$\alpha$ forest region (below 8,400 $\rm \AA$).} 
\label{fig:spec}
\end{figure*}

\subsection{Temporal and spectral evolution during the Swift observation}
\indent 
{\it Swift}/XRT  observations showed a high hydrogen column density decreasing with time, which can be interpreted as a time-dependent photoionization of the local circumburst medium, within a compact and dense environment, as found in only a few cases.

The XRT light curve (Fig.~\ref{fig:lc}) reveals noticeable variations in the observed count rate that can be divided into 4 segments. The results from the time resolved spectral analysis from the X-ray emission of GRB\,130606A are given in Table~\ref{table:xrtlog}. The dependence of the flux on frequency, $\nu$, and time t, is described through this section and in Table~\ref{table:xrtlog} by $F_\nu$ $\propto$ $\mathrm{\nu^{-\beta}}$ $\mathrm{t^{-\alpha}}$ where $\beta$ =  $\Gamma$ – 1. Thus, we find the following distinct episodes:

I) Beginning of XRT observations up to $\sim$ $T_0$+759 s. This segment is dominated by flaring activity from internal shocks as part of the prompt emission. The combined {\it Swift}/BAT-XRT X-ray spectrum in the interval $T_0$ + 155s to $T_0$ + 175s can be fit using a simple absorbed power law model ($\mathrm{\chi^{2}}$/d.o.f. = 1.29), and the addition of a thermal component has negligible effect ($\mathrm{\chi^{2}}$/d.o.f. = 1.26). The best fit model (Fig.~\ref{fig:bat+xrtspec}) has a hydrogen column density (4 $\pm$ 2) $\times$ $\mathrm{10^{22} \,cm^{-2}}$ and a photon index $\Gamma$ = 1.03 $\pm$ 0.02.

II) $T_0$+759 s to $T_0$+1300 s. This is characterised by a fast decay with temporal index, $\alpha_X$=3 $\pm$ 1, consistent with high-latitude emission \citep{Genet:2009aa} at the end of the prompt emission ($\alpha_X$= 2+$\beta_X$), where $\alpha_X$ = 0.62 $\pm$ 0.09.

III) $T_0$+1300 s to $T_0$+1.6 $\times$ $\mathrm{10^{4}}$ s. During this segment, the X-ray count-rate decay is consistent with a temporal decay index of $\alpha_X$ =0.66 $\pm$ 0.20. This plateau phase is typically associated with late activity from the central engine \citep{Zhang:2007aa}. The end of this plateau phase at $T_0$+1.6 $\times$ $\mathrm{10^{4}}$ s seems to be achromatic when comparing the X-ray and i-band light curves (Fig.~\ref{fig:lc}) as expected at the end of late activity from the central engine. The closure relations modified with an energy injection parameter, $q$ \citep{Zhang:2006aa}, are consistent with an homogeneous environment (ISM) with $q$ = 0.5 $\pm$ 0.4 when $\nu_x < \nu_c$. This value is consistent with previous measurements of $q$ \citep{Zhang:2006aa, Curran:2009aa}. The electron spectral index inferred during this segment is $p$ = 2.84 $\pm$ 0.30, consistent within 1$\sigma$ with the distribution of values of $p$ presented by \cite{Curran:2010aa} and \cite{Starling:2008aa}. The wind model when $\nu_x < \nu_c$  with late energy injection is rejected for GRB\,130606A since the $q$-parameter inferred is $q<0$ which does not have physical meaning. The cases when $\nu_x > \nu_c$  in the ISM or wind model are also rejected since result in $p<2$ for which the closure relations are no longer valid. Finally, the optical temporal decay index $\alpha_{opt}$ $\sim$ 1.3 is consistent with  $\alpha_{X}$ during this time interval.

IV)  $T_0$+1.6 $\times$ $\mathrm{10^{4}}$ s to $T_0$+3 $\times$ $\mathrm{10^{5}}$s. In this segment, the decay ($\alpha_X$ =1.86 $\pm$ 0.20) is consistent with forward shock emission when $\nu_x < \nu_c$. The closure relation between the temporal and spectral index in the case of an ISM model for $\nu_x < \nu_c$ is $\alpha$ = 3$\beta$/2 = 1.55 $\pm$ 0.30 consistent with the observed $\alpha_X$. For a wind environment, the relationship can be re-written as $\alpha$=(3$\beta$+1)/2=2.03 $\pm$ 0.30 which is also consistent with the observed temporal decay. However, the wind environment was tentatively rejected in the previous segment and may not be necessary to explain the afterglow emission of GRB\,130606A. The electron spectral index obtained in this segment is, p = 3.0 $\pm$ 0.4. This value is consistent with the electron spectral index inferred in the previous segment. It should be noted that during this segment there seems to be small variability in the X-ray light curve in the form of a micro-flare peaking at $ \sim T_0$+4 $\times$ $\mathrm{10^{4}}$ s. As shown in Fig.~\ref{fig:xrtlc} GRB\,130606A shows significant variation in the photon index, $\Gamma$ and $N_H$ column density throughout the X-ray observations. The variation of  $\Gamma$ with respect to the X-ray luminosity shows that the spectrum gets progressively softer as the flux decreases.

The intrinsic column density $N_H$ is well constrained while the central engine remains active (first 500 s since trigger) implying high levels of photoionization of the local high-density medium. During this time, the observed intrinsic $N_H$  is almost 3 orders of magnitude higher than the galactic column density at these coordinates. Once the prompt emission ends, the intrinsic $N_H$  decreases abruptly and no excess with respect to the galactic value can be found for the remainder of the X-ray observations. This can be interpreted as a time-dependent photoionization of the local circumburst medium, within a compact and dense environment, only found in a few GRBs such as GRB\,980329 \citep{Lazzati:2002aa} and GRB\,000528 \citep{Frontera:2004aa}.

\begin{figure}
\centering
   \includegraphics[width=9cm]{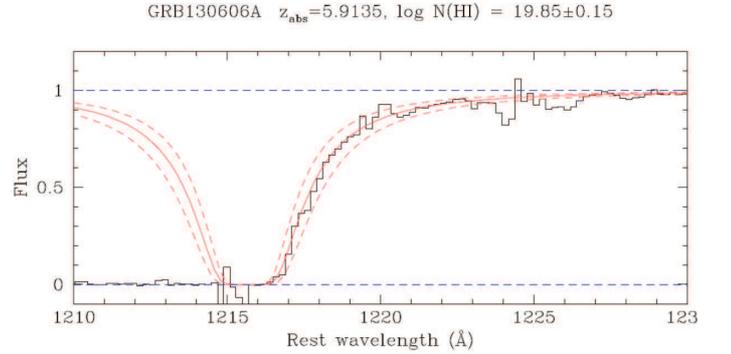}
\caption{The N (H I) fit to the GTC (+ OSIRIS) spectrum of GRB\,130606A. Taken on June 7, 2013, the figure shows the data (black solid line) and the best fit damped profile (solid red line).  The derived column density is log N (H I) = 19.85, together with the fits for log N (H I) = 19.70 and 20.00 (dashed red lines).} 
\label{fig:nh-fit}
\end{figure} 

\subsection{The GRB\,130606A sub-DLA host galaxy}
\indent 
In order to determine the neutral hydrogen content at the GRB host galaxy redshift, the Lyman-alpha line observed in the GTC optical spectrum (Fig.~\ref{fig:spec}) was fitted with a damped profile and a value of log N (H I) = 19.85 $\pm$ 0.15 was derived (Fig.~\ref{fig:nh-fit}), in good agreement with two independent data-sets \citep{Chornock:2013aa,Totani:2013aa}.  The associated system is therefore technically classified as a sub-DLA.

Besides structure in the Lyman forest region (discussed in \cite{Chornock:2013aa}), the GTC optical spectrum shows a variety of absorption lines at different redshifts (e.g. 2.310, 2.521, 3.451), but we will concentrate on the GRB\,130606A system at z = 5.9135 that is associated with the host galaxy. 
%The list of detected absorption lines are displayed in Table XX."

%In order to derive the elemental abundances, various methods were applied to determine robust column densities: a basic integrated equivalent width measurement, application of the apparent optical depth method, and gaussian line fitting.  A summary of the measurements is given in Table~\ref{table:metlog}. 

The high signal-to-noise ratio (SNR) of the GTC spectrum permitted  a search for relatively weak metal lines, and offers an improvement over some of the limits measured by \cite{Chornock:2013aa}.
%for S, Si, O and C.  
%
%[S/H] < -0.82;  [Si/H] = -1.80 $\pm$ 0.15;  [O/H] = -2.06 $\pm$ 0.16 and  [C/H] = -1.78 $\pm$ 0.16, which includes a contribution from CII.  The limit of the sulphur abundance, determined from the non-detection of the SII 1253 \AA line, is a factor of two deeper than that found previously \cite{Chornock:2013aa}, thanks to the high SNR of our GTC spectrum.  
%

In the case of sulphur, the triplet at 1250, 1253, 1259 \AA~ can be used. Significant absorption is detected at the position of the weakest of these three lines (1250 \AA), but the lack of absorption at 1253 \AA~ indicates that the absorption is likely from a contaminating source. Based on the non-detection of the S II  1253 \AA~ line, we determine an observed frame 3 $\sigma$ EW limit $< 0.157$ \AA~ (assuming a FWHM = 3.4 \AA~ and a S/N of 65 in the S II line region) which corresponds to  0.023 \AA~ in the rest frame.  The rest-frame EW limit yields log N(S II) $<$ 14.17.  Assuming N(H I) = 19.85 and solar S/H = 4.85 (from \cite{Asplund:2009aa}) this gives a 3$\sigma$ upper limit [S/H] $<$ -0.82, which is 0.3 dex (a factor of $\sim$ 2) deeper than the sulphur limit obtained by \cite{Chornock:2013aa}.

In addition to the upper limit to the sulphur abundance, we can determine lower
limits to the abundances of oxygen and silicon.  The limiting silicon abundance is determined from the mildly saturated Si II 1260 \AA~ line with a rest frame EW=0.35 \AA, yielding [Si/H] $>$ -1.80, without consideration of ionization or dust depletion corrections.  The oxygen abundance is determined from the O I 1302 \AA~ line, which is also likely to be partly saturated despite its modest EW (0.2 \AA). The fact that O I doesn't require an ionization correction and O does not deplete, means that this is one of the best lines from which to obtain an accurate metallicity. The major uncertainty here is that it is close to a small noise feature on the red side which might lead to an over-estimate of the oxygen abundance at the 0.1 dex level. Taking these factors into account, we determine [O/H]$> -$2.06.  The Si and O limits are consistent to within 0.1 dex of the values independently derived (from different spectra) by \cite{Chornock:2013aa}.  Combined with the upper limit from sulphur, we can constrain the metallicity within a factor of about 10, in the range from $\sim$1/7 to $\sim$1/60 of solar.

%, which places the system in a metallicity regime where saturation of O I 1302 \AA~ and C II 1334 \AA~ are expected. Therefore only lower limits on the total column densities for both C and O can be obtained (i.e. it is not possible to determine the C/O ratio). Same applies for Si II 1260 \AA, which is highly saturated, as also shown by \cite{Chornock:2013aa}.
%
%Besides the above mentioned saturation issues, 
For a more comprehensive study of the abundances, we refer 
to \cite{Hartog:2013aa}. 
Furthermore, we also point out that it is very 
likely that the gas is partially ionized: strong high-ionization lines 
(such as Si IV and N V) are present at the redshift of the absorber. 
%Even if the O I arises entirely from neutral gas, there may be significant 
%C II associated with the high-ionization phase, what will obviously increase 
%the total C II 1334 \AA~ EW. 
%implying an increase of the implied C/O ratio.  

 %__________________________________________________     
%\begin{table}
%\caption{Column densities and abundances for spectroscopic data obtained at the 10.4m GTC (z=5.9135)} 
%\label{table:metlog}
%\centering
%\begin{tabular}{ccc}
%\hline
%Element      &  Log N(X) &  [X/H]        \\
%\hline
%H	&    19.85 $\pm$ 0.15 &	   ...   \\     
%O	&   <14.48            & < -2.06  \\
%C	&   <14.50            & < -1.78  \\
%S       &   <14.17            & < -0.82  \\
%Si	&   <13.56            & < -1.80  \\
%\hline
%\end{tabular}
%\end{table}
%__________________________________________________ 
\begin{table*}
\caption{Spectral fitting analysis for the {\it Swift}/XRT X-ray data assuming
N$_{H}$ (Gal) = 2 $\times$ 10$^{20}$ cm$^{-2}$.} 
\label{table:xrtlog}
\begin{tabular}{ccccc}
\hline
\noalign{\smallskip}
Time Interval (after $T_0$ [s] ) &  $N_H(intrinsic)$ [$\mathrm{10^{22} \,cm^{-2}}$] &   $\Gamma$  & $F_X (unabsorbed)$  & $\mathrm{\chi^{2}}$/dof \\
\noalign{\smallskip}
\hline
\noalign{\smallskip}
             
78-89	 &     10(-5.,+6)  &  1.10(-0.12,+0.13) &   3.25e-09(-0.24,+0.22) &  42/43\\
115-200  &  4.1(-1.0,+1.1) & 1.24(-0.04,+0.04) &  2.10e-09(-0.06,+0.06) & 241/225\\
200-300 &   8.3(-1.0,+1.0) & 1.60(-0.03,+0.03) &  2.61e-09(-0.05,+0.05) & 324/296\\
300-500  &  3.5(-0.6,+0.7) & 1.81(-0.04,+0.04) &  8.88e-10(-0.22,+0.25) & 285/244\\
\hline
78-200	&     5.1(-1.0,+1.1) & 1.22(-0.04,+0.04) & 2.85e-09(-0.08,+0.09) &  261/259\\
\hline
500-1000  &   $<$ 1.4    & 	 1.62(-0.09,+0.09)  & 1.07e-10(-0.12,+0.05)  & 24/28\\
1000-1e4  &   $<$  3 	 &       1.92(-0.14,+0.15) & 8.4e-12(-0.7,+0.7)  &     26/26\\
1e4-3e4  &  $<$ 1.1  &1.87(-0.14,+0.15) &  2.7e-12(-0.3,+0.3)   &    27/27\\
 3e4-3e5	&   $<$  1  &   2.03(-0.13,+0.20) & 2.9e-13(-0.4,+0.2) & 10/13\\
500-3e5	& $<$ 0.5 &  1.81(-0.06,+0.07) & 1.81e-12(-0.07,+0.08) &   65/89\\
\noalign{\smallskip}
\hline
\end{tabular}
\end{table*}
    %__________________________________________________ One column table

\begin{figure}
\centering
   \includegraphics[angle=-90,width=9cm]{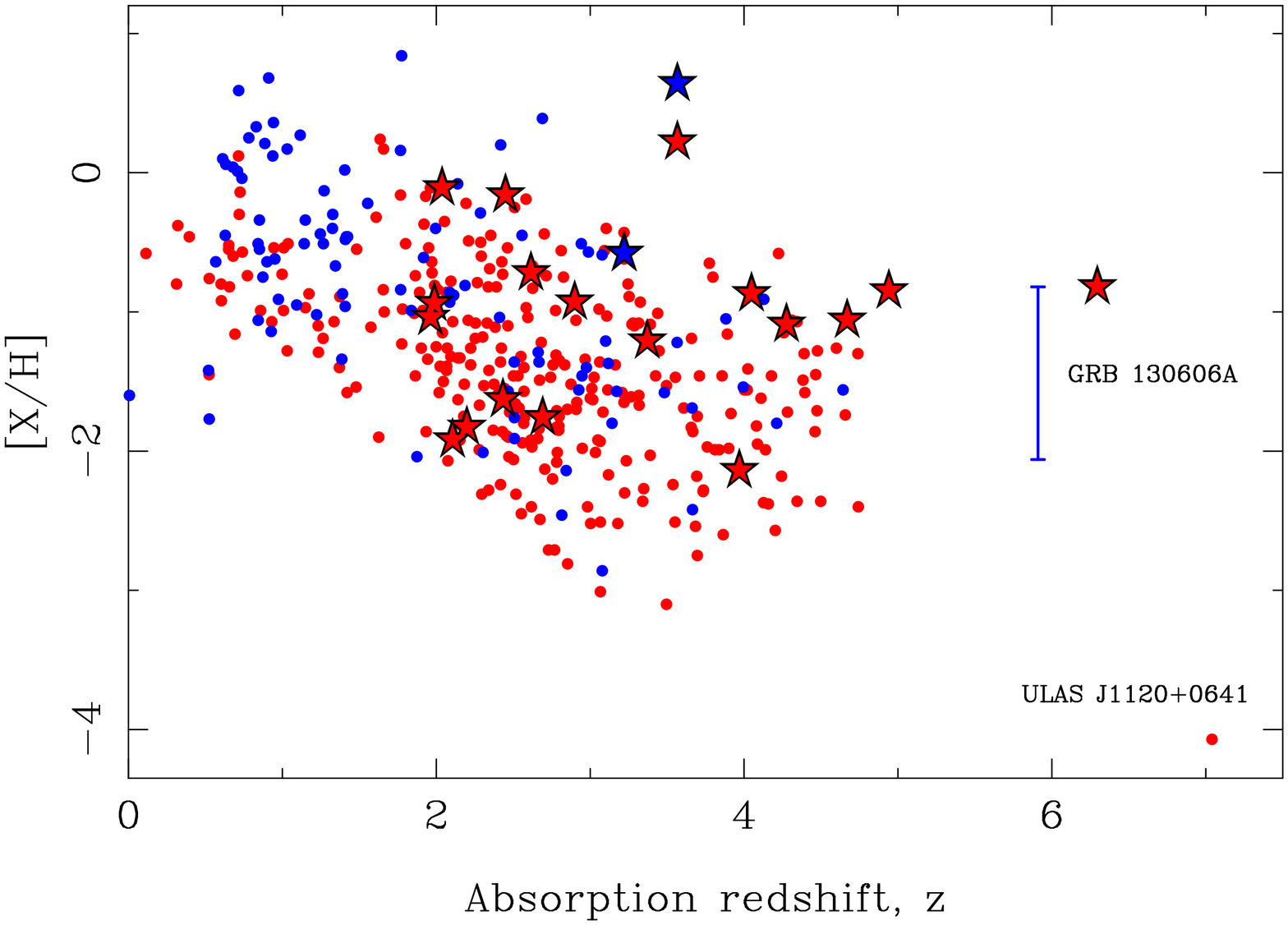}
\caption{The metallicity ([X/H]) as a function of redshift is shown for a compilation of QSO-DLAs (circles)
%,  \cite{Berg:2013aa}) 
and GRB-DLAs (stars, \cite{Schady:2011aa,Thone:2013aa}), including the location for GRB 130606A at z = 5.9 (blue error bar) and ULAS J1120+0641 at z $\sim$ 7 \citep{Simcoe:aa}. The GRB 130606A sub-DLA is the 2nd highest redshift burst with a measured GRB-DLA metallicity and only the third GRB absorber with sub-DLA HI column density. Blue colours are used for log N(HI) $<$ 20.3 and red is used for log N(H I) $\geq$ 20.3.  In order of preference for any given absorber, Zn, S, O, Si, Fe+0.4 dex is our choice of metallicity indicator, where the 0.4 offset for Fe accounts for typical dust depletion.} 
\label{fig:metal-z}
\end{figure} 
   
\begin{figure}
\centering
   \includegraphics[angle=-90,width=9cm]{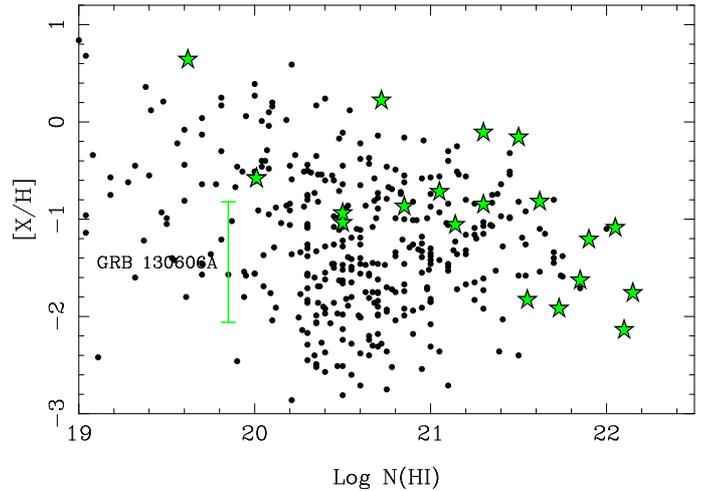}
\caption{The metallicity of a GRB sample (green stars) versus N(H I) compared to a sample of DLAs (black dots). Adapted from \cite{Schady:2011aa} and \cite{Thone:2013aa}. The location for GRB\,130606A (green error bar) is also plotted.} 
\label{fig:metal-nh}
\end{figure}

\section{Conclusions}
\label{Conclusions}
With an initial Lorentz bulk factor in the range $\Gamma_0$ $\sim$ 65-220, the X-ray afterglow evolution can be explained by a time-dependent photoionization of the local circumburst medium, within a compact and dense environment. The host galaxy is a sub-DLA (log N (H I) = 19.85 $\pm$ 0.15), and a metallicity 
content in the range from $\sim$1/7 to $\sim$1/60 of solar is inferred.

In order to place the chemistry of the GRB sub-DLA in context with other high z absorbers, both Fig.~\ref{fig:metal-z} and Fig.~\ref{fig:metal-nh} show the metallicity of a compilation of GRB host galaxy absorption systems (GRB-DLAs) compared to quasars with DLA and sub-DLAs (QSO-DLAs)
%, \cite{Berg:2013aa})
, combining the data reported in the literature \citep{Schady:2011aa, Thone:2013aa}.  The GRB\,130606A sub-DLA is a rare find: the second highest redshift burst with a measured GRB-DLA metallicity and only the third GRB absorber with sub-DLA HI column density.
%, and by far the lowest metallicity of the three. In fact, only one other GRB has a metallicity comparably low. 
At z $>$ 5, the only other object known with lower metallicity is the ULAS J1120+0641 DLA at z $\sim$ 7 \citep{Simcoe:aa}. However, the DLA towards ULAS J1120+0641 is close to the redshift of the quasar and its metallicity is determined from a stacked spectrum, both of which complicate its interpretation \citep{Ellison:2010aa, Ellison:2011aa}. 
%The host for GRB\,130606A harbours the only z $>$ 5 DLA system with a direct metallicity measurement below 1/100 solar.

%It is relatively rare to be able to determine abundances of carbon and oxygen in DLAs,  because the principal lines tend to be saturated.  However, at low metallicities (and/or low N(HI)) this problem is considerably mitigated \citep{Cooke:2011ab}.  By combining the abundances of carbon and oxygen stated above, we determine a relative abundance of [C/O] = +0.28 $\pm$ 0.05. 
%(see Fig~\ref{ffig:xrtlc}). 
%
%A high value of C/O is predicted to be a signature of Pop III stellar enrichment \citep{Fabbian:2009aa}.  A handful of DLAs in the metallicity range -2 to -3 (1/100 to 1/1000 of the Sun's metallicity) have C/O measurements \citep{Ellison:2010aa, Cooke:2011ab} and all but one (towards the quasar QSO J0035-0918) have [C/O] < 0 \citep{Cooke:2011ab}. Whilst the GRB\,130606A system is not as C-enhanced as the DLA in QSO J0035-0918 at z = 2.3 (whose [C/O] = 0.77), it is a factor of 2 above the solar ratio and considerably higher than all of the  other DLA and stellar points for that metallicity \citep{Cooke:2011ab}. Possible corrections to the observed C/O ratio are discussed in the SI, but we propose the  supersolar C/O at the GRB host galaxy redshift is the nucleosynthetic signature of Pop III stars in a such a system at z = 5.9.

We note that GRB\,130606A, given the non-zero metal content of the host, might have originated from a non-Pop III progenitor star, but whether its afterglow light penetrated material that was pre-enriched by Pop III nucleosynthesis at even higher redshifts \citep{Wang:2012aa} remains uncertain. Indeed several possibilities for the death of the first stars have recently been suggested by theoretical models \citep{Bromm:2013aa}.

Events such as GRB\,130606A at z = 5.91, and future ones at z $>$ 10, offer an exciting new window into pre-galactic metal enrichment in these very high redshift galaxies. These bright lighthouses constitute a significant step forward towards using these sources as beacons for measuring abundances at such early times.

New GRB missions, equipped with on-board near-IR detectors, and coupled to state-of-the-art instruments built for the largest diameter ground-based telescopes, such as the forthcoming CIRCE instrument \citep{Eikenberry:2013aa} on the 10.4m GTC, will allow us to study the first stars that fundamentally transformed the Universe only a few hundred million years after the Big Bang.

\begin{acknowledgements}
Partly based on observations carried out with the 0.6 m TELMA telescope at the BOOTES-2 station in EELM-CSIC, with the 10.4 m Gran Telescopio Canarias instaled in the Spanish Observatorio del Roque de los Muchachos of the Instituto de Astrof\'{\i}sica de Canarias in the island of La Palma (GTC69-13A) and with the IRAM Plateau de Bure Interferometer. IRAM is supported by INSU/CNRS (France), MPG (Germany) and IGN (Spain). We acknowledge the support of F. J. Aceituno (OSN observatory) and of the Spain’s Ministerio de Ciencia y Tecnolog\'{\i}a through Projects AYA2009-14000-C03-01/ESP, AYA2011-29517-C03-01 and AYA2012-39727-C03-01 and Creative Research Initiatives of MEST/NRF in Korea (RCMST). R.H. acknowledges GA CR grant 102/09/0997. E.S. acknowledge support by the Scientific and Technological Research Council of Turkey (T\"UBITAK) through the project 112T224. A.P. and A.V. are partially supported by RFBR 12-02-01336, 13-01-92204. LH, AMC and MT acknowledge support from SFI under grants 07/RFP/PHYF295 and 09/RFP/AST/2400 and from the EU FP7 under grant agreement 283783. T.G. acknowledges support from Bilim Akademisi - The Science Academy, Turkey, under the BAGEP program. This study has been developed in the framework of the Unidad Asociada IAA/CSIC-UPV/EHU, supported  by the Ikerbasque science foundation. J.G. is partially supported by AYA2012-39362-C02-02. We acknowledge the use of public data from the {\it Swift} data archives and the service provide by the gamma-ray burst Coordinates Network (GRB) and BACODINE system, maintained by S. Barthelmy. 
\end{acknowledgements}

\bibliographystyle{aa}

\begin{thebibliography}{41}
\expandafter\ifx\csname natexlab\endcsname\relax\def\natexlab#1{#1}\fi


\bibitem[{Afonso {et~al.}(2013)Afonso, Kann, Nicuesa, Kruehler, Elliot \& Greiner}]{Afonso:2013aa}
Afonso, P., Kann, D. A., Nicuesa, A. {et~al.} 2013, GRB Coordinates Network, 14807, 1

\bibitem[{Asplund {et~al.}(2009)Asplund, Grevesse, Sauval, \&
  Scott}]{Asplund:2009aa}
Asplund, M., Grevesse, N., Sauval, J.~A., \& Scott, P. 2009, ARAA, 47, 481

\bibitem[{Barthelmy {et~al.}(2013)Barthelmy, Baumgartner, Cummings, Fenimore,
  Gehrels, Krimm, \& Lien}]{Barthelmy_gcn2013}
Barthelmy, S.~D., Baumgartner, W.~H., Cummings, J.~R., {et~al.} 2013, GRB
  Coordinates Network, 14819, 1

\bibitem[{Berg {et~al.}(2013)}]{Berg:2013aa}
Berg, T. A. {et~al.} 2013, in preparation

\bibitem[{Bromm(2013)}]{Bromm:2013aa}
Bromm, V. 2013, RPPh, 76, 2901

\bibitem[{Bromm \& Loeb(2002)}]{Bromm:2002aa}
Bromm, V. \& Loeb, A. 2002, ApJ, 575, 111

\bibitem[{Bromm \& Loeb(2006)}]{Bromm:2006aa}
Bromm, V. \& Loeb, A. 2006, ApJ, 642, 382

\bibitem[{Bromm {et~al.}(2009)Bromm, Yoshida, Hernquist, \& McKee}]{Bromm:aa}
Bromm, V., Yoshida, N., Hernquist, L., \& McKee, C.~F. 2009, Nature, 459, 49

\bibitem[{Butler {et~al.}(2013)Butler, Watson, Kutyrev et al.}]{Butler:2013aa}
Butler, N., Watson, A. M., Kutyrev, A. {et~al.} 2013, GRB Coordinates Network, 14824, 1

\bibitem[{Castro-Tirado {et~al.}(2013{\natexlab{a}})Castro-Tirado,
  Sanchez-Ramirez, Jelinek, Gorosabel, Tello, Ferrero, Lara-Gil, \&
  Cunniffe}]{Castro-Tirado:2013aa}
Castro-Tirado, A.~J., Sanchez-Ramirez, R., Jelinek, M., {et~al.}
  2013{\natexlab{a}}, GRB Coordinates Network, 14790, 1

\bibitem[{Castro-Tirado {et~al.}(2013{\natexlab{b}})Castro-Tirado,
  Sanchez-Ramirez, Gorosabel, Jelinek, Tello, Ferrero, Lara-Gil, Cunniffe,
  Perez-Ramirez, \& Kubanek}]{Castro-Tirado:2013ab}
Castro-Tirado, A.~J., Sanchez-Ramirez, R., Gorosabel, J., {et~al.}
  2013{\natexlab{b}}, GRB Coordinates Network, 14796, 1

\bibitem[{Chornock {et~al.}(2013)Chornock, Berger, Fox, Lunnan, Drout, Fong, \&
  Laskar}]{Chornock:2013aa}
Chornock, R., Berger, E., Fox, D.~B., {et~al.} 2013, ApJ, 774, 26

\bibitem[{Ciardi \& Loeb(2000)}]{Ciardi:2000aa}
Ciardi, B. \& Loeb, A. 2000, ApJ, 540, 687

\bibitem[{Cooke {et~al.}(2011{\natexlab{a}})Cooke, Pettini, Steidel, Rudie, \&
  Jorgenson}]{Cooke:2011aa}
Cooke, R., Pettini, M., Steidel, C.~C., Rudie, G.~C., \& Jorgenson, R.~A.
  2011{\natexlab{a}}, Monthly Notices of the Royal Astronomical Society, 412,
  1047

\bibitem[{Cooke {et~al.}(2011{\natexlab{b}})Cooke, Pettini, Steidel, Rudie, \&
  Nissen}]{Cooke:2011ab}
Cooke, R., Pettini, M., Steidel, C.~C., Rudie, G.~C., \& Nissen, P.~E.
  2011{\natexlab{b}}, MNRAS, 417, 1534

\bibitem[{Curran {et~al.}(2010)Curran, Evans, de~Pasquale, Page, \& van~der
  Horst}]{Curran:2010aa}
Curran, P.~A., Evans, P.~A., de~Pasquale, M., Page, M.~J., \& van~der Horst,
  A.~J. 2010, ApJ, 716, L135

\bibitem[{Curran {et~al.}(2009)Curran, Starling, van~der Horst, \&
  Wijers}]{Curran:2009aa}
Curran, P.~A., Starling, R. L.~C., van~der Horst, A.~J., \& Wijers, R. A. M.~J.
  2009, Mon. Not. R. Astron. Soc, 395, 580

\bibitem[{Eikenberry {et~al.}(2013)Eikenberry, Lasso, Raines, Charcos, Edwards,
  Garner, Mar{\'\i}n-Franch, Cenarro, Bennett, \&
  Frommeyer}]{Eikenberry:2013aa}
Eikenberry, S.~S., Lasso, N., Raines, S.~N., {et~al.} 2013, RMxAC, 42, 119

\bibitem[{Ellison {et~al.}(2011)Ellison, Prochaska, \& Mendel}]{Ellison:2011aa}
Ellison, S., Prochaska, J.~X., \& Mendel, J.~T. 2011, MNRAS, 412, 448

\bibitem[{Ellison {et~al.}(2010)Ellison, Prochaska, Hennawi, Lopez, Usher,
  Wolfe, Russell, \& Benn}]{Ellison:2010aa}
Ellison, S.~L., Prochaska, J.~X., Hennawi, J., {et~al.} 2010, MNRAS, 406, 1435

\bibitem[{Fabbian {et~al.}(2009)Fabbian, Nissen, Asplund, Pettini, \&
  Akerman}]{Fabbian:2009aa}
Fabbian, D., Nissen, P.~E., Asplund, M., Pettini, M., \& Akerman, C. 2009,
  A\&A, 500, 1143

\bibitem[{Frontera {et~al.}(2004)Frontera, Amati, Lazzati, Montanari,
  Orlandini, Perna, Costa, Feroci, Guidorzi, Kuulkers, Masetti, Nicastro,
  Palazzi, Pian, \& Piro}]{Frontera:2004aa}
Frontera, F., Amati, L., Lazzati, D., {et~al.} 2004, ApJ, 614, 301

\bibitem[{Genet and Granot (2009)Genet \& Granot}]{Genet:2009aa}
Genet, F. and Granot, J. 2009, MNRAS, 399, 1328

\bibitem[{Ghirlanda {et~al.}(2012)Ghirlanda, Nava, Ghisellini, Celotti, Burlon,
  Covino, \& Melandri}]{Ghirlanda:2012aa}
Ghirlanda, G., Nava, L., Ghisellini, G., {et~al.} 2012, MNRAS, 420, 483

\bibitem[{Golenetskii {et~al.}(2013)Golenetskii, Aptekar, Mazets, Pal'Shin,
  Frederiks, Oleynik, Ulanov, Svinkin, \& Cline}]{Golenetskii:2013aa}
Golenetskii, S., Aptekar, R., Mazets, E., {et~al.} 2013, GRB Coordinates
  Network, 14808, 1

\bibitem[{Guilloteau {et~al.}(1992)Guilloteau, Delannoy, Downes, Greve, Guelin, Lucas, Morris, Radford, Wink, Cernicharo}]{Guilloteau:1992aa}
Guilloteau, S., Delannoy, J., Downes, D. {et~al.} 1992, A\&A, 262, 624

\bibitem[{Hartog {et~al.}(2013)Hartog et al.}]{Hartog:2013aa}
Hartog, O. {et~al.} 2013, in preparation

\bibitem[{Jelinek {et~al.}(2013)Jelinek, Gorosabel, Castro-Tirado, Mottola,
  Hellmich, Fernandez-Munoz, \& Munoz-Martinez}]{Jelinek:2013aa}
Jelinek, M., Gorosabel, J., Castro-Tirado, A.~J., {et~al.} 2013, GRB
  Coordinates Network, 14782, 1

\bibitem[{Karlsson {et~al.}(2013)Karlsson, Bromm, \&
  Bland-Hawthorn}]{Karlsson:2013aa}
Karlsson, T., Bromm, V., \& Bland-Hawthorn, J. 2013, RvMP, 85, 809

\bibitem[{Lamb \& Reichart(2000)}]{Lamb:2000aa}
Lamb, D.~Q. \& Reichart, D.~E. 2000, ApJ, 536, L1

\bibitem[{Laskar {et~al.}(2013)Laskar, Zauderer, \& Berger}]{Laskar:2013aa}
Laskar, T., Zauderer, A., \& Berger, E. 2013, GRB Coordinates Network, 14817, 1

\bibitem[{Lazzati(2002)}]{Lazzati:2002aa}
Lazzati, D. \&~Perna, R. 2002, MNRAS, 330, 383

\bibitem[{Liang {et~al.}(2010)Liang, Yi, Zhang, L{\"u}, Zhang, \&
  Zhang}]{Liang:2010aa}
Liang, E.-W., Yi, S.-X., Zhang, J., {et~al.} 2010, ApJ, 725, 2209

\bibitem[{L{\"u} {et~al.}(2012)L{\"u}, Zou, Lei, Zhang, Wu, Wang, Liang, \&
  L{\"u}}]{Lu:2012aa}
L{\"u}, J., Zou, Y.-C., Lei, W.-H., {et~al.} 2012, ApJ, 751, 49

\bibitem[{Mackey {et~al.}(2003)Mackey, Bromm, \& Hernquist}]{Mackey:2003aa}
Mackey, J., Bromm, V., \& Hernquist, L. 2003, ApJ, 586, 1

\bibitem[{Melandri {et~al.}(2010)Melandri, Kobayashi, Mundell, Guidorzi,
  de~Ugarte~Postigo, Pooley, Yoshida, Bersier, Castro-Tirado, Jel{\'\i}nek, \&
  Gomboc}]{Melandri:2010aa}
Melandri, A., Kobayashi, S., Mundell, C.~G., {et~al.} 2010, ApJ, 723, 1331

\bibitem[{Panaitescu and Vestrand(2011)Panaitescu and Vestrand}]{Panaitescu:2011aa}  
Panaitescu A. \& Vestrand, W. T. 2011, MNRAS, 414, 3537

\bibitem[{Persson {et~al.}(1998)Persson et al.}]{Person:1998aa}  
Persson S. E., Murphy, D. C., Krzeminski, W., {et~al.} 1998, AJ, 116, 2475

\bibitem[{Rykoff {et~al.}(2009)Rykoff, Aharonian, Akerlof, Ashley, Barthelmy,
  Flewelling, Gehrels, \& G{\"o}{\v g}{\"u}{\c s}}]{Rykoff:2009aa}
Rykoff, E.~S., Aharonian, F., Akerlof, C.~W., {et~al.} 2009, ApJ, 702, 489

\bibitem[{Scannapieco {et~al.}(2005)Scannapieco, Madau, Woosley, Heger, \&
  Ferrara}]{Scannapieco:2005aa}
Scannapieco, E., Madau, P., Woosley, S., Heger, A., \& Ferrara, A. 2005, ApJ,
  633, 1031

\bibitem[{Schady {et~al.}(2011)Schady, Savaglio, Kr{\"u}hler, Greiner, \&
  Rau}]{Schady:2011aa}
Schady, P., Savaglio, S., Kr{\"u}hler, T., Greiner, J., \& Rau, A. 2011, A\&A,
  525, A113

\bibitem[{Simcoe {et~al.}(2012)Simcoe, Sullivan, Cooksey, Kao, Matejek, \&
  Burgasser}]{Simcoe:aa}
Simcoe, R.~A., Sullivan, P.~W., Cooksey, K.~L., {et~al.} 2012, Nature, 492, 79

\bibitem[{Starling {et~al.}(2008)Starling, van~der Horst, Rol, Wijers,
  Kouveliotou, Wiersema, Curran, \& Weltevrede}]{Starling:2008aa}
Starling, R. L.~C., van~der Horst, A.~J., Rol, E., {et~al.} 2008, ApJ, 672, 433

\bibitem[{Th{\"o}ne {et~al.}(2013)Th{\"o}ne, Fynbo, Goldoni, Postigo~de Ugarte,
  Campana, Vergani, \& Covino}]{Thone:2013aa}
Th{\"o}ne, C.~C., Fynbo, J. P.~U., Goldoni, P., {et~al.} 2013, MNRAS, 428, 3590

\bibitem[{Totani {et~al.}(2013)Totani, Aoki, Hattori, Kosugi, Niino, Hashimoto,
Kawai, Ohta, Sakamoto \& Yamada}]{Totani:2013aa}
Totani, T., Aoki, K., Hattori, T. {et~al.} 2013, PASJ, submitted (arXiv:1312.3934v1)

\bibitem[{Ukwatta {et~al.}(2013)Ukwatta, Barthelmy, Gehrels, Krimm, Malesani,
  Marshall, Maselli, \& Melandri}]{Ukwatta:2013aa}
Ukwatta, T.~N., Barthelmy, S.~D., Gehrels, N., {et~al.} 2013, GRB Coordinates
  Network, 14781, 1

\bibitem[{Virgili {et~al.}(2013)Virgili, Mundell, Melandri \& Gomboc}]{Virgili:2013aa}
Virgili, F. J., Mundell, C., Melandri, A. \& Gomboc, A. 2013, GRB Coordinates Network, 14785, 1

\bibitem[{Wang {et~al.}(2012)Wang, Bromm, Greif, Stacy, Dai, Loeb, \&
  Cheng}]{Wang:2012aa}
Wang, F.~Y., Bromm, V., Greif, T.~H., {et~al.} 2012, ApJ, 760, 27

\bibitem[{Wolfe {et~al.}(2005)Wolfe, Gawiser, \& Prochaska}]{Wolfe:aa}
Wolfe, A.~M., Gawiser, E., \& Prochaska, J.~X. 2005, ARAA, 43, 861

\bibitem[{Zhang(2007)}]{Zhang:2007aa}
Zhang, B. 2007, CJAA, 7, 1

\bibitem[{Zhang {et~al.}(2006)Zhang, Fan, Dyks, Kobayashi, M{\'e}sz{\'a}ros,
  Burrows, Nousek, \& Gehrels}]{Zhang:2006aa}
Zhang, B., Fan, Y.~Z., Dyks, J., {et~al.} 2006, ApJ, 642, 354

\end{thebibliography}

\end{document}